\documentclass[aps,prl,twocolumn,showpacs,floats]{revtex4}
\usepackage{latexsym}
\usepackage{amsmath}
\usepackage{amssymb}
\usepackage{bm}
\usepackage{wasysym}
\usepackage[dvips]{color}
\usepackage{graphicx}
\usepackage{graphicx}
\usepackage{subfigure}
\usepackage{epsfig}
\usepackage{ifpdf}

\newcommand{\veps}{\varepsilon}

\def\g{\mathrm{g}}

\def\e{\mathrm{e}}

\def\mbfk{\mathbf{k}}

\def\mbfp{\mathbf{p}}

\def\mbfr{\mathbf{r}}
\def\mbfR{\mathbf{R}}

\def \mbfA{\mathbf{A}}

\def\3P0{$^{3}$P$_{0}$}
\def\1S0{$^{1}$S$_{0}$}

\usepackage{amsmath,stmaryrd,graphicx}
\makeatletter
\newcommand{\fixed@sra}{$\vrule height 2\fontdimen22\textfont2 width 0pt\shortrightarrow$}
\newcommand{\shortarrow}[1]{%
  \mathrel{\text{\rotatebox[origin=c]{\numexpr#1*45}{\fixed@sra}}}
}
\makeatother

\begin{document}

\title{Optical Control of Exchange Interaction and Kondo Temperature in cold Atom Gas}

\author{Igor Kuzmenko$^1$, Tanya Kuzmenko$^1$ and Yshai Avishai$^{1,2,3}$}

\affiliation{$^1$Department of Physics,
  Ben-Gurion University of the Negev,
  Beer-Sheva, Israel \\
  $^2$New York University and the NYU-ECNU Institute
  of Physics at NYU Shanghai, 3663 Zhongshan Road North,
  Shanghai, 200062, China \\
  $^3$Yukawa Institute for Theoretical Physics, Kyoto, Japan}

\begin{abstract}
  The relevance of magnetic impurity problems
  in cold atom systems depends crucially on
  the nature of the exchange interaction between
  itinerant fermionic atoms and localized
  impurity atoms. In particular, Kondo physics
  occurs only if the exchange interaction is
  anti-ferromagnetic, and strong enough to
  yield high enough Kondo temperature
  ($T_K/T_F \ge 0.1$). Focusing, as an example,
  on the experimentally accessible system
  of ultra-cold $^{173}$Yb atoms, it is shown
  that the sign and strength of an exchange
  interaction between an itinerant Yb(\1S0)
  atom and a trapped Yb(\3P0) atom can be
  optically controlled. Explicitly, as the light
  intensity increases (from zero),
  the exchange interaction changes from
  ferromagnetic to anti-ferromagnetic.
  When the light intensity is such that the system approaches
  a singlet Feshbach resonance (from below), the singlet
  scattering length $a_S$ is large and
  negative, and the Kondo temperature
  increases sharply.
\end{abstract}
\date{\today}

\pacs{37.10.Jk, 31.15.vn, 33.15.Kr}

\maketitle

\noindent
{\underline{\textbf{Introduction}}}:
Controlling interaction between cold atoms is
a godsend, as it turns atomic systems capable of
demonstrating new phenomena that cannot be
otherwise accessed within
solid state physics proper
\cite{exch-PRL-09,exch-RevModPhys-13,%
exch-Nature-98,exch-PRL-15,exch-RevModPhys-10}.
In a series of theoretical
\cite{laser-exch-PRL-96,laser-control-exch-PRL-2003,%
opt-tune-scatt-length-PRA-05,opt-Feshbach-PRA-2006}
and experimental \cite{Feshbach-RevModPhys-2006,%
laser-exch-PRL04,laser-exch-PRA97,laser-exch-PRL11,%
laser-exch-Bachelors-2009,opt-Feshbach-prl-11,laser-exch-Science-2008}
investigations, it has been established that,
as far as {\it  potential scattering} is concerned,
 the strength of
atomic interactions can be tuned by laser beams. The prime object
of these studies is to achieve an optical Feshbach resonance and
thereby to obtain a Bose-Einstein condensate in
cold bosonic atom systems.

On the other hand, the feasibility of controlling the strength and the sign
of {\it exchange interaction}  between atoms is much less studied.
Its importance has been recognized recently, in the quest for studying the
Kondo effect \cite{Hewson-book} in cold atom systems \cite{Bauer}.
The physics exposed in the study of magnetic impurities
when the itinerant (fermionic) atoms have spin $F\geq\frac{3}{2}$
is very rich, touching upon exotic phenomena such as over-screening
\cite{KKAK-prb-15}, realization of the Coqblin-Schrieffer
model \cite{IK-TK-YA-GBJ-PRB-16}, multipolar
Kondo effect \cite{IK-TK-YA-GBJ-PRB-18} and others.
Recently, it has been shown that exchange interaction can be
controlled by the technique of confined-induced-resonance (CIR)
\cite{CIR1,CIR2,CIR3,CIR4}.

The goal of the present study is to
show that exchange interaction
between fermionic atoms can be {\it optically controlled}
\footnote{Comparison between the CIR and optical techniques
will be assessed elsewhere}.
As an experimentally feasible example we consider
$^{173}$Yb atoms. Those in
the ground state \1S0  are  itinerant, (forming a degenerate Fermi gas confined in a
shallow square well potential), whereas those in the long lived excited
state \3P0 are trapped in a state-dependent optical potential
and serve as dilute concentration of
localized magnetic impurities. Both the ground \1S0 and excited \3P0 state
atoms have spin $F=\frac{5}{2}$ (that is
the nuclear spin).
In this case, the Coqblin-Schrieffer model
with SU(6) symmetry \cite{Hewson-book} is realizable due to
a unique exchange mechanism
\cite{IK-TK-YA-GBJ-PRB-16}.
 It is shown that by applying
 laser beams on the atomic gas, the exchange
interaction between Yb(\1S0) and Yb(\3P0)
atoms can be controlled, both in sign and magnitude.
In particular, with increasing
the light intensity, the exchange interaction
changes from ferromagnetic to
anti-ferromagnetic,  that is a pre-requisite for occurrence of the Kondo physics.

\begin{figure}[htb]
\centering
  \includegraphics[width=70 mm,angle=0]
   {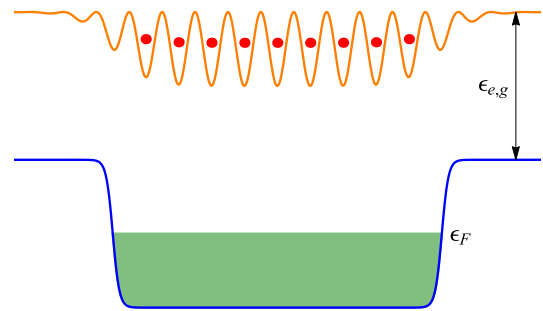}
 \caption{{\color{blue}(Color online)}
  Illustration of a magnetic impurity setup for $^{173}$Yb atoms (schematic).
   Atoms in the ground-state \1S0 form a Fermi sea
   (Fermi energy $\epsilon_F$), while atoms in
   the excited-state \3P0 are trapped in an optical
   potential and form a dilute concentration of localized
   magnetic impurities.
   $\epsilon_{\e,\g}$ is the excitation energy
   (\ref{energy-eg}) of the \3P0 state.}
 \label{Fig-Vg-Ve}
\end{figure}

\noindent
{\underline{\textbf{Description of the system}}}:
Consider a 3D gas of $^{173}$Yb atoms confined in
a shallow square well potential.
Most of the atoms remain in the ground state
(\1S0) and form a Fermi sea due to its half integer
nuclear spin $I=\frac{5}{2}$ (see green area in
Fig. \ref{Fig-Vg-Ve}).
However,  following a coherent excitation
via the clock transition, a few atoms occupy the long-lived
excited state (electronic configuration \3P0). These excited atoms can
be trapped in a state-dependent optical lattice
potential as schematically displayed in
Fig. \ref{Fig-Vg-Ve} (red circles), and
can be regarded as dilute concentration of localized impurities.

\vspace{-0.02in}

The energy dispersion and density  of states (DOS) for the (weakly confined)
Yb(\1S0) atoms read,
\begin{eqnarray}
  \epsilon_k ~=~
  \frac{\hbar^2 k^2}{2 M},
  \ \
  \rho(\epsilon)
  ~=~
  \frac{M k_{\epsilon}}{2 \pi^2 \hbar^2}.
  \label{disp}
\end{eqnarray}
Here $M$ is the atomic mass, $k=|\mbfk|$, $\mbfk=(k_x,k_y,k_z)$, with
$k_{\alpha=x,y,z}=\frac{\pi n_{\alpha}}{2 L}$, in which $\{n_\alpha \}$ are
integers and $k_{\epsilon}=\sqrt{2 M \epsilon}/\hbar$.

As for the impurities, an Yb(\3P0) atom at position
${\bf R}=(X_1,X_2,X_3)$ is trapped by an optical potential,
\begin{eqnarray}
  V_{\e}(\mbfR) &=&
  -V_{\e}^{(0)}
  \sum_{\alpha}
  \sin^2\big(k_{\e} X_{\alpha}\big),
  \label{Ve-def}
\end{eqnarray}
where $k_{\e}$ is the wave number of the laser light.
The lowest energy level $\epsilon_{\mathrm{imp}}$ of
the Yb(\3P0) atom is deep and close to the minimum of
$V_{\e}(\mbfR)$,  hence it can be
approximated harmonic potential at lowest energy,
\begin{eqnarray}
  \epsilon_{\mathrm{imp}} &=&
  \frac{3}{2}~
  \hbar \Omega_{\e},
  \label{e-imp}
\end{eqnarray}
wherein the harmonic frequency and length are,
\begin{eqnarray}
  \hbar \Omega_{\e} ~=~
  2
  \sqrt{{\mathcal{E}}_{\e}
        V_{\e}^{(0)}},
  \ \ \ \ \
  k_{\e} a_{\e} ~=~
  \bigg(
       \frac{{\mathcal{E}}_{\e}}{V_{\e}^{(0)}}
  \bigg)^{1/4},
  \label{harmonic-length-frequency}
\end{eqnarray}
with recoil energy
$
  {\mathcal{E}}_{\e} =
  \frac{\hbar^2 k_{\e}^{2}}{2 M}.
$
\ \\
\noindent
{\underline{\textbf{Exchange interaction}}
between the Yb(\1S0) and
Yb(\3P0) atoms is
described in details in
Ref. \cite{IK-TK-YA-GBJ-PRB-16}, but there,
one of the parameters is chosen wrongly, leading to an incorrect conclusion. Here we
correct it  and explain
how the exchange interaction can be optically controlled.
Consider an Yb atom as a doubly ionized
closed shell rigid ion and two valence electrons.
The ground state \1S0 electron configuration
is $6s^2$, whereas that of the excited state
\3P0 is $6s6p$.
The excitation energy is
\begin{eqnarray}
  \epsilon_{\e,\g} &=&
\epsilon_{\e}-\epsilon_{\g}=  2.14349~{\text{eV}}.
  \label{energy-eg}
\end{eqnarray}
The positions of the ion core and the outer
electrons are respectively specified by
vectors $\mbfR$, $\mbfr_{\mathrm{a}}$
and $\mbfr_{\mathrm{b}}$.
In the adiabatic (Born-Oppenheimer) approximation
(which is well substantiated in atomic physics),
the wave function of a single Yb atom is
expressed as a product of the wave functions
of the rigid ion core (considered as a point particle of
mass $M$), and that of the valence electrons.
 When one valence electron virtually tunnels from
an Yb($6s^2$) atom to an Yb($6s6p$) atom we get an ionized
Yb$^{+}$($6s$) atom and a charged
Yb$^{-}$($6s^26p$) atom.
The corresponding excitation energy is,
\begin{eqnarray}
  \Delta\veps &=&
  \veps_{\mathrm{ion}}-
  \veps_{\mathrm{ea}}-
  \epsilon_{\e,\g}
  ~=~
  4.4107~{\mathrm{eV}},
  \label{Delta-veps}
\end{eqnarray}
where $\veps_{\mathrm{ion}}=6.2542$~eV is the ionization
energy of ytterbium \cite{Yb-spectrum-78} and
$\veps_{\mathrm{ea}}=-0.3$~{eV} is the electron
affinity \cite{Electron-afinity-PhRep04}.

Such virtual tunneling of electrons between the atoms
gives rise to anti-ferromagnetic exchange
interaction between them.
Following Ref.~\cite{IK-TK-YA-GBJ-PRB-16},
we describe the exchange interaction by the potential
\begin{eqnarray}
  V_{\mathrm{exch}}^{\mathrm{(bare)}}(R) =
  g_0 \zeta(R),
  \label{V-exch-bare}
\end{eqnarray}
where
\begin{eqnarray*}
  g_0 &=&
  2.821472~
  {\text{eV}} \times {\text{\AA}}^{3},
  \\
  \zeta(R) &=&
  \frac{1}{Z}~
  \bigg(\frac{R}{r_0}\bigg)^{4 \gamma}
  e^{-\kappa (R-r_0)},
  \\
  Z &=&
  \frac{4 \pi r_{0}^{3}}{\big(\kappa r_0\big)^{4 \gamma + 3}}~
  \Gamma\big(4 \gamma + 3,\kappa r_0\big)~
  e^{\kappa r_0}.
\end{eqnarray*}
Here $r_0$ is the classical turning point where the van der Waals
potential vanishes [see Eq.~ (\ref{vdW-def}) below].
The parameters $\kappa$ and $\gamma$ are
\begin{eqnarray*}
  &&
  \kappa =
  \kappa_s +
  \kappa_p =
  2.3199~{\text{\AA}}^{-1},
  \\
  &&
  \gamma =
  \frac{1-\beta_s}{\beta_s} +
  \frac{1-\beta_p}{\beta_p} =
  1.2942,
\end{eqnarray*}
where $\beta_{\mathrm{s}}=0.67799$,
$\beta_{\mathrm{p}}=0.54966$,
$\kappa_{\mathrm{s}}=1.2812$~{\AA}$^{-1}$ and
$\kappa_{\mathrm{p}}=1.0387$~{\AA}$^{-1}$
\cite{IK-TK-YA-GBJ-PRB-16}.

\noindent
{\underline{\textbf{Van der Waals interaction}}}:
The van der Waals interaction between the Yb atoms is modelled here
by the Lennard-Jones potential
\cite{vdW-Yb-PRA-08,vdW-Yb-PRA-14},
\begin{eqnarray}
  W^{\mathrm{(bare)}}(R) &=&
  \frac{C_6}{R^6}~
  \bigg\{
       \frac{\sigma^6}{R^6}-1
  \bigg\}-
  \frac{C_8}{R^8},
  \label{vdW-def}
\end{eqnarray}
with $C_6=2.561\cdot10^3E_ha_B^6$,
$C_8=3.222465\cdot10^5E_ha_B^8$ and
$\sigma=9.0109361a_B$, where
$E_h=27.211$~eV is the Hartree energy and
$a_B=0.52918$~{\AA} is the Bohr radius.

\noindent
{\underline{\textbf{Light-assisted interaction}}}:
Controlling the strength and sign of the exchange interaction is achieved by
subjecting the mixture of Yb atoms to a laser beam
of frequency
$\omega_0$ that is tuned to be close to the resonant
frequency $\omega_{\mathrm{res}}=\Delta\veps/\hbar$.
Recall that $\Delta\veps$ is the energy difference (\ref{Delta-veps})
between the two neutral atoms and the two ions. Concretely,
$
  \frac{\omega_{\mathrm{res}}}
       {2 \pi c}
  ~=~
  35574.7~{\text{cm}}^{-1}.
$

The interaction between the electromagnetic field
(derived from a vector potential ${\bf A}$) and electrons in an atom is of the form \cite{Landau-Lifshitz-2},
\begin{eqnarray*}
  H_{\mathrm{em}} &=&
  -\frac{e}{c}~
  \frac{\mbfp \cdot \mbfA}{m_e},
\end{eqnarray*}
where $\mbfp=-i \hbar \nabla$ is the electronic momentum
operator and $m_e$ is the electronic mass. This interaction
gives rise to tunneling of electrons between the atoms.
When detuning $\Delta\omega=\omega_0-\omega_{\mathrm{res}}$
of the light frequency is much larger than the natural linewidth
of the absorption line $\omega_{\mathrm{res}}$, tunnelling of electrons
between the atoms is forbidden by the energy conservation law.
In second order perturbation theory, the laser light induces
an additional potential $W^{\mathrm{(ind)}}(R)$ and
exchange interactions $V_{\mathrm{exch}}^{\mathrm{(ind)}}(R)$
between the neutral atoms,
\begin{eqnarray}
  W^{\mathrm{(ind)}}(R)
  &=&
  \frac{1}{\hbar \Delta \omega}~
  \Big\{
      t_{s,{\mathrm{ind}}}^{2}(R)+
      t_{p,{\mathrm{ind}}}^{2}(R)
  \Big\},
  \label{W-exch-induced}
  \\
  V_{\mathrm{exch}}^{\mathrm{(ind)}}(R)
  &=&
  \frac{2}{\hbar \Delta \omega}~
  t_{s,{\mathrm{ind}}}(R)~
  t_{p,{\mathrm{ind}}}(R).
  \label{V-exch-induced}
\end{eqnarray}
The tunneling rates $t_{\mu,{\mathrm{ind}}}(R)$ ($\mu=s,p$) are,
\begin{eqnarray*}
  t_{\mu,{\mathrm{ind}}}(R)
  &=&
  \frac{e v_0 E_0}{4 \pi \omega_0}~
  {\mathfrak{F}}_{\mu}(R),
\end{eqnarray*}
where $v_0$=$\frac{\hbar}{m_e a_B}$=$0.00729735c$,
$E_0$ is the amplitude of the laser's electric field and
the dimensionless functions ${\mathcal{F}}_{\mu}(R)$ are
\begin{eqnarray}
  {\mathfrak{F}}_{\mu}(R)
  =
  a_B
  \int d^3\mbfr~
  \psi_{\mu}^{*}\big(\big|\mbfr-\mbfR\big|\big)~
  \frac{\partial
        \psi_{\mu}(r)}
       {\partial z}.
  \label{Is-Ip-WF-single-electron}
\end{eqnarray}
The axis $z$ is chosen parallel to the vector
$\mbfR=\mbfR_{\g}-\mbfR_{\e}$.
For the single-electron wave function $\psi_{\mu}(r)$, we use
the asymptotic expression \cite{IK-TK-YA-GBJ-PRB-16},
\begin{eqnarray}
  \psi_{\mu}(r) &=&
  \sqrt{\frac{2 {\mathcal{A}}~\kappa_{\mu}^{3}}
             {\pi~
              \Gamma
              \big(
                  \frac{\beta_{\mu}+2}
                       {\beta_{\mu}}
              \big)}}~
  \big(
      2 \kappa_{\mu} r
  \Big)^{\frac{1-\beta_{\mu}}{\beta_{\mu}}}
  e^{-\kappa_{\mu} r},
  \label{WF-single-electron}
\end{eqnarray}
where $\mu=s,p$ for the $6s$ and $6p$ electron,
${\mathcal{A}}=1.355$,
\begin{eqnarray*}
  &&
  \beta_{\mathrm{s}} ~=~
  0.67799,
  \ \ \ \ \
  \kappa_{\mathrm{s}} ~=~
  1.2812~{\text{\AA}}^{-1},
  \\
  &&
  \beta_{\mathrm{p}} ~=~
  0.54966,
  \ \ \ \ \
  \kappa_{\mathrm{p}} ~=~
  1.0387~{\text{\AA}}^{-1}.
\end{eqnarray*}
It is useful to rewrite the (laser induced) potential and exchange
contributions [$W^{\mathrm{(ind)}}(R)$ and
$V_{\mathrm{exch}}^{\mathrm{(ind)}}(R)$] in
the compact form,
\begin{eqnarray}
  W^{\mathrm{(ind)}}(R) &=&
  \frac{V_0}{2}~
  \Big\{
      {\mathcal{F}}_{s}^{2}(R)+
      {\mathcal{F}}_{p}^{2}(R)
  \Big\},
  \label{W-ind-res}
  \\
  V_{\mathrm{exch}}^{\mathrm{(ind)}}(R) &=&
  V_0~
  {\mathcal{F}}_{s}(R)~
  {\mathcal{F}}_{p}(R),
  \label{V-exch-ind-res}
\end{eqnarray}
where the coupling $V_0$ is given by,
\begin{eqnarray}
  &&
  V_0 ~=~
  \frac{2}{\hbar \Delta \omega}~
  \bigg(
       \frac{e E_0 v_0}{4 \pi \omega_0}
  \bigg)^{2}.
  \label{V0-def}
\end{eqnarray}
Hereafter we consider the case of blue detuning with
$\Delta \omega > 0$.

 \noindent
{\underline{\textbf{Scattering lengths}}}:
The van der Waals and exchange interactions
yield ``singlet'' and ``triplet'' scattering lengths
\cite{Scatt-Length-WKB-PRA93,vdW-Yb-PRA-08,%
IK-TK-YA-GBJ-PRB-16},
\begin{eqnarray}
  a_{\nu} &=&
  \bar{a}~
  \bigg\{
       1-
       \tan
       \Big(
           \Phi_{\nu}-
           \frac{\pi}{8}
       \Big)
  \bigg\},
  \label{scattering-length}
\end{eqnarray}
where $\nu=S,T$ for the quantum states with antisymmetric
(``singlet'') and symmetric (``triplet'') two-particle spin wave
functions, and
\begin{eqnarray}
  \bar{a} &=&
  \frac{1}{2^{3/2}}~
  \frac{\Gamma\big(\frac{3}{4}\big)}
       {\Gamma\big(\frac{5}{4}\big)}~
  \bigg(
       \frac{M C_6}{\hbar^2}
  \bigg)^{1/4}
  ~=~
  39.73~{\text{\AA}}.
  \label{bar-a}
\end{eqnarray}
The parameters $\Phi_{\nu}$ are,
\begin{eqnarray}
  \Phi_{\nu} &=&
  \int\limits_{R_{\nu}}^{\infty}
  K_{\nu}(R)~dR,
  \label{Phi-nu}
\end{eqnarray}
where
\begin{eqnarray}
  K_{\nu}(R) &=&
  \frac{1}{\hbar}~
  \sqrt{-M
        \big[
            W(R)+
            \eta_{\nu}
            V_{\mathrm{exch}}(R)
        \big]}.
  \label{K-def}
\end{eqnarray}
In the above equation $W(R) =W^{\mathrm{(bare)}}(R)+%
W^{\mathrm{(ind)}}(R)$,
$V_{\mathrm{exch}}(R)=%
V_{\mathrm{exch}}^{\mathrm{(bare)}}(R)+%
V_{\mathrm{exch}}^{\mathrm{(ind)}}(R)$,
$\eta_S=1$ and $\eta_T=-1$.
$R_{\nu}$ is the solution of equation
$W(R_{\nu})+\eta_{\nu}V_{\mathrm{exch}}(R_{\nu})=0$.

When the intensity of the laser beam with frequency
$\omega_0$ vanishes [i.e., when $V_0\to0$],
the scattering lengths are \cite{IK-TK-YA-GBJ-PRB-16}
\begin{eqnarray}
  a_S ~=~
  1878~a_B,
  \ \ \ \ \
  a_T ~=~
  219.7~a_B,
  \label{scattering-length-E0=0-num}
\end{eqnarray}
(where $a_B$ is the Boh'r radius)
which agree well with the experimental results
\cite{Scazza-Yb-3P0,Scazza-Yb-3P0-2015}.

Evaluating the scattering lengths for nonzero
$V_0$, requires calculation of
$V_{\mathrm{exch}}^{\mathrm{(ind)}}(R)$ and
$W^{\mathrm{ind}}(R)$, eqs. (\ref{V-exch-ind-res})
and (\ref{W-ind-res}).
Substituting
them into eq. (\ref{Phi-nu})
yields the quantity $\Phi_{\nu=S,T}$
as functions of $V_0$.
This is carried out numerically,
after which eq. (\ref{scattering-length}) is employed
to find the scattering lengths.

\begin{figure}[htb]
\centering
  \includegraphics[width=80 mm,angle=0]
   {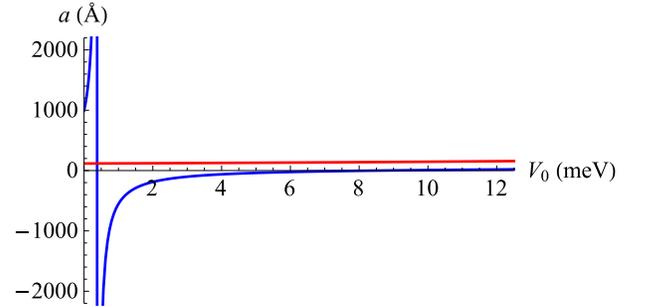}
 \caption{{\color{blue}(Color online)}
   Scattering lengths $a_S$ [blue curve]
   and $a_T$ [red curve] as functions
   of $V_0$.}
 \label{Fig-scatt-length-vs-V0}
\end{figure}
\vspace{-0.2in}
The scattering lengths (\ref{scattering-length})
calculated numerically are displayed in
Fig. \ref{Fig-scatt-length-vs-V0} as functions of $V_0$.
It is seen that in the interval $V_0<12.5$~{meV}, there
is a value of $V_0$ where $a_S$ is singular:
$V_{{\mathrm{c}}}=0.3809$~{meV}.
On  the other hand, $a_T$ varies slowly within
the interval $V_0<12.5$~meV}.

\noindent
{\underline{\textbf{Kondo Hamiltonian}}}:
Once the ``singlet'' and ``triplet'' scattering lengths are known,
it is possible to construct the effective Kondo Hamiltonian,
with explicit expressions for potential and exchange coupling constants
denoted below as $ g_{\mathrm{pot}}$ and $g_{\mathrm{exch}}$ respectively.
Let us consider a degenerate Fermi gas of Yb(\1S0)
atoms, and one Yb(\3P0) atom localized at $\mbfR_{\e}=0$
which plays a role of a magnetic impurity. Then the Kondo
Hamiltonian is,
\begin{eqnarray}
  H_K &=&
  g_{\mathrm{pot}}
  \sum_{\mbfk,\mbfk'}
  \sum_{m}
  c_{\mbfk',m}^{\dag}
  c_{\mbfk,m}+
  \nonumber \\ &+&
  g_{\mathrm{exch}}
  \sum_{\mbfk,\mbfk'}
  \sum_{m \neq m'}
  X^{m,m'}
  c_{\mbfk',m'}^{\dag}
  c_{\mbfk,m}+
  \nonumber \\ &+&
  g_{\mathrm{exch}}
  \sum_{\mbfk,\mbfk'}
  \sum_{m}
  Z^{m,m}
  c_{\mbfk',m}^{\dag}
  c_{\mbfk,m}.
  \label{HK-def}
\end{eqnarray}
Here $c_{\mbfk,m}$ and $c_{\mbfk,m}^{\dag}$ are
annihilation and creation operators of itinerant atom
with wave vector $\mbfk$ and magnetic quantum
number $m$. $X^{m,m'}=|{m}\rangle\langle{m'}|$
are Hubbard operators of the localized impurity,
the ket $|m\rangle$ describes the impurity
with magnetic quantum number $m$.
$Z^{m,m}=X^{m,m}-\frac{1}{6}$. $g_{\mathrm{pot}}$
and $g_{\mathrm{exch}}$ are effective couplings
of the potential and exchange interactions.
Here $g_{\mathrm{exch}}>0$ denotes antiferromagnetic
exchange interaction.
The laser induced scattering lengths due to the short range interaction
(\ref{HK-def}) are,
\begin{eqnarray}
  a_{\nu=S,T} &=&
  \frac{M}{4 \pi \hbar^2}~
  \big\{
       g_{\mathrm{pot}}+
       \alpha_{\nu}~
       g_{\mathrm{exch}}
  \big\},
  \label{scatt-lengths-vs-couplings}
\end{eqnarray}
where $\alpha_S=-\frac{7}{6}$ and $\alpha_T=\frac{5}{6}$.
Then the couplings $g_{\mathrm{pot}}$ and $g_{\mathrm{exch}}$
of the potential and exchange interactions are expressed
is terms of the scattering lengths as,
\begin{eqnarray}
  &&
  g_{\mathrm{pot}} ~=~
  \frac{\pi \hbar^2}{3 M}~
  \Big\{
      5 a_S+7 a_T
  \Big\},
  \label{g-pot-vs-scatt-lengths}
  \\
  &&
  g_{\mathrm{exch}} ~=~
  \frac{2 \pi \hbar^2}{M}~
  \Big\{
      a_T-a_S
  \Big\}.
  \label{g-exch-vs-scatt-lengths}
\end{eqnarray}
Eq. (\ref{g-exch-vs-scatt-lengths}) shows that
when $a_T>a_S$, the exchange interaction is
anti-ferromagnetic and the
Hamiltonian (\ref{HK-def})
gives rise to Kondo effect. When $a_T<a_S$,
the exchange interaction is ferromagnetic and
there is no Kondo effect.

\noindent
{\underline{\textbf{Kondo temperature}}}:
When the exchange interaction (\ref{HK-def}) is
anti-ferromagnetic (corresponding to regions in Fig.~\ref{Fig-scatt-length-vs-V0}
 where $a_T>a_S$), the Kondo temperature
is \cite{Hewson-book},
\begin{eqnarray}
  T_K &=&
  D_0~
  \exp
  \bigg(
       -\frac{1}{6~g_{\mathrm{exch}}~\rho(\epsilon_F)}
  \bigg).
  \label{TK-res}
\end{eqnarray}
Hereafter we assume that $2 \hbar \Omega_{\e}<\epsilon_F$
[where $\epsilon_F$ is the Fermi energy and
the harmonic frequency $\Omega_{\e}$ is given
by eq. (\ref{harmonic-length-frequency})],
and therefore $D_0=2 \hbar \Omega_{\e}/k_B$
plays a role of ultraviolet cutoff of the Kondo theory.
The constant $\rho(\epsilon_F)$
is the DOS (\ref{disp}) at the Fermi energy.
A simple  expression relating $T_K$ to the scattering lengths
(valid for $a_T>a_S$) then reads,
\begin{eqnarray}
  T_K &=&
  D_0~
  \exp
  \Bigg(
       -\frac{\pi}{6 k_F \big(a_T-a_S\big)}
  \Bigg),
  \label{TK-vs-scatt-lengths}
\end{eqnarray}
where $a_T$ and $a_S$ as functions of $V_0$ are
shown in Fig. \ref{Fig-scatt-length-vs-V0}.

\begin{figure}[htb]
\centering
  \includegraphics[width=80 mm,angle=0]
   {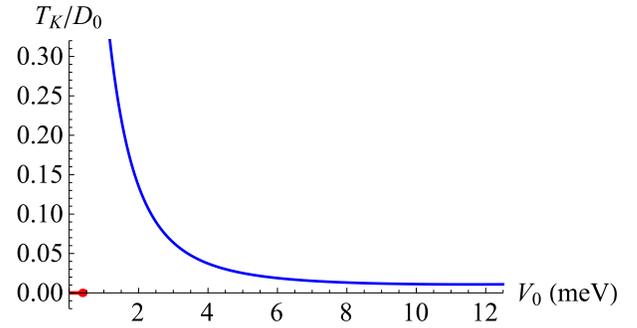}
   \vspace{-0.15in}
 \caption{{\color{blue}(Color online)}
   Kondo temperature (blue curves) (\ref{TK-vs-scatt-lengths})
   as a function of $V_0$ for $\epsilon_F/k_B = 100$~nK.
   The red segment indicates the interval  $V_0<V_{\mathrm{c}}$
   for which the Kondo effect is absent.
   Thus, the red dot denotes the critical potential $V_{{\mathrm{c}}}$ which separates
   the interval where  Kondo effect is present from that
   where it is absent.}
 \label{Fig-TK-vs-V0}
 \vspace{-0.1in}
\end{figure}
The Kondo temperature (\ref{TK-vs-scatt-lengths})
calculated numerically is shown in Fig. \ref{Fig-TK-vs-V0},
(blue curves). It is seen that the Kondo temperature
sharply increases with decreasing $a_S$ [see
Fig. \ref{Fig-scatt-length-vs-V0}]. In the red segment of
$V_0$ where $a_S>a_T$ the Kondo
effect is absent.

\noindent
{\underline{\textbf{Conclusions}}}:
Tuning interaction strength in quantum impurity
problems in cold atom systems cannot rely on
the application of an external magnetic field
(for driving Feshbach resonance),
because it is detrimental for the Kondo effect.
Hence, employing optical toolbox for controlling interaction
strength in cold atom systems
is a proper substitute. But so far it has been
applied in numerous works mainly for studying bosonic
systems. The quest for studying quantum (magnetic) impurity
problems and Kondo physics requires a novel kind of {\it controlling
 the exchange interaction}.
More concretely, it involves a subtle tuning of ``singlet'' and
``triplet'' scattering lengths, and identifying
the conditions wherein $a_S<a_T$, for which
the exchange interaction is antiferromagnetic.

This objective has been achieved here for
an experimentally representative system.
The feasibility to construct
optically tuneable exchange interaction
between itinerant $^{173}$Yb(\1S0)
atoms and a localized $^{173}$Yb(\3P0)
impurity has been analyzed. ``Singlet''
and ``triplet'' scattering lengths as
a function of $V_0$ (which is proportional
to the light intensity) are explicitly calculated
and the regions where $a_S<a_T$ in which
the exchange interaction is antiferromagnetic
are identified. With increasing
intensity of light (from zero), the exchange
interaction changes both in magnitude and in sign.
The Kondo Hamiltonian is then constructed
and the Kondo temperature is calculated
in the intervals of $V_0$ where the exchange
interaction is  anti-ferromagnetic. It is shown
that $T_K$ increases sharply before reaching
an optical Feshbach resonance where
the singlet scattering length approaches
$-\infty$ \footnote{From the present
analysis we now conclude that the claim that
antiferromagnetic exchange exists {\it  also in the absence of laser field}
as reported in Ref. \cite{IK-TK-YA-GBJ-PRB-16} should be retracted.
Fortunately, it can be
curred following the analysis detailed in the present note. A proper
corrigendum will shortly be reported.}.
\ \\
\noindent
\underline{Acknowledgement}
We thank the authors of Refs.\cite{CIR1,CIR2,CIR3,CIR4} for drawing our attention
to the CIR method. Prof. Y. Band is to be acknowledged for pointing out the
relevance of spontaneous emission.
This research is supported in part  by an Israel Science Foundation grant 400/12.

\newpage

\subsection{About spontaneous emission}

Spontaneous emission is dangerous since
it heats the atomic gas. In order to avoid this,
we need to organize the procedure as follows:

(1) The frequency of detuning from the resonant
frequency is needed to be large enough  so that
the decay rate due to spontaneous emission of
a photon is very low.

(2) The \1S0 and \3P0 quantum states have
different electronic principal quantum numbers
and spin states. Electric dipole transition
between the singlet states \1S0 and \3P0 is forbidden.
Magnetic dipole transition
can change the spin, but not the principal quantum
number. Therefore, spontaneous quantum
transition from \3P0 to \1S0 state of Yb is
virtually forbidden (or at least, the lifetime of the \3P0
state is very long).

Note that the same problem exists also for  a "bare"
mixture of Yb(\1S0) and Yb(\3P0) atoms
(without an additional laser light).
However, applying an additional laser radiation makes
the exchange interaction stronger and
(possible) the lifetime shorter. So far, we do not
know how to calculate the lifetime.

In an experimental work arXiv:1708.03810 \cite{CIR3}
the authors write:
``For the scattering of the ionization beam, we will
try to estimate the scattering rates and compare
it with our trap depths. Probably, we will have
to choose a significant detuning as not to heat
the atoms out of the trapping confinement.''


\begin{thebibliography}{99}

\bibitem{exch-PRL-09} S. E. Pollack, D. Dries, M. Junker,
         Y. P. Chen, T. A. Corcovilos, and R. G. Hulet,
         Phys. Rev. Lett. {\textbf{102}}, 090402 (2009).

\bibitem{exch-RevModPhys-13} Dan M. Stamper-Kurn
         and Masahito Ueda,
         Rev. Mod. Phys. {\textbf{85}}, 1191 (2013).

\bibitem{exch-Nature-98} J. Stenger, S. Inouye,
         D. M. Stamper-Kurn, H.-J. Miesner, A. P. Chikkatur and
         W. Ketterle, Nature {\textbf{396}}, 345 (1998).

\bibitem{exch-PRL-15} Simon Murmann, Andrea Bergschneider,
         Vincent M. Klinkhamer, Gerhard Z\"urn, Thomas Lompe,
         and Selim Jochim, Phys. Rev. Lett. {\textbf{114}}, 080402 (2015).

\bibitem{exch-RevModPhys-10} Cheng Chin, Rudolf Grimm,
         Paul Julienne, and Eite Tiesinga,
         Rev. Mod. Phys. {\textbf{82}}, 1225 (2010).


\bibitem{laser-exch-PRL-96} P. O. Fedichev, Yu. Kagan,
         G. V. Shlyapnikov, and J. T. M. Walraven,
         Influence of nearly resonant light on the scattering
         length in low-temperature atomic gases,
         Phys. Rev. Lett. {\textbf{77}}, 2913 (1996).

\bibitem{laser-control-exch-PRL-2003} L.-M. Duan,
         E. Demler, M. D. Lukin,
         Phys. Rev. Lett. {\textbf{91}}, 090402 (2003);
         arXiv:cond-mat/0210564.

\bibitem{opt-tune-scatt-length-PRA-05} R. Ciury{\l}o,
         E. Tiesinga, and P. S. Julienne,
         Phys. Rev. A {\textbf{71}}, 030701(R) (2005).

\bibitem{opt-Feshbach-PRA-2006} Pascal Naidon and
         Fran\c{c}oise Masnou-Seeuws,
         Phys. Rev. A {\textbf{73}}, 043611 (2006).


\bibitem{Feshbach-RevModPhys-2006} Thorsten K\"ohler,
         Krzysztof G\'oral, and Paul S. Julienne,
         Rev. Mod. Phys. {\textbf{78}}, 1311 (2006).

\bibitem{laser-exch-PRL04} M. Theis, G. Thalhammer,
         K. Winkler, M. Hellwig, G. Ruff, R. Grimm,
         J. Hecker Denschlag,
         Tuning the scattering length with an optically induced
         Feshbach resonance,
         Phys. Rev. Lett. {\textbf{93}}, 123001 (2004);
         arXiv:cond-mat/0404514.

\bibitem{laser-exch-PRA97} John L. Bohn and
         P. S. Julienne, Prospects for influencing scattering
         lengths with far-off-resonant light,
         Phys. Rev. A {\textbf{56}}, 1486 (1997).

\bibitem{laser-exch-PRL11} Felix H.J. Hall, Mireille Aymar,
         Nadia Bouloufa-Maafa, Olivier Dulieu, Stefan Willitsch,
         Light-assisted ion-neutral reactive processes in
         the cold regime: radiative molecule formation vs.
         charge exchange,
         Phys. Rev. Lett. {\textbf{107}}, 243202 (2011);
         arXiv:1108.3739.


\bibitem{laser-exch-Bachelors-2009} Adam Micah Kaufman,
         Radiofrequency dressing of atomic Feshbach resonances,
         Submitted to the Department of Physics of Amherst College
         in partial fulfilment of the requirements for the degree of
         Bachelors of Arts, 2009.

\bibitem{opt-Feshbach-prl-11} S. Blatt, T. L. Nicholson, B. J. Bloom,
         J. R. Williams, J. W. Thomsen, P. S. Julienne, and J. Ye,
         Phys. Rev. Lett. {\textbf{107}}, 073202 (2011).

\bibitem{laser-exch-Science-2008} S. Trotzky, P. Cheinet,
         S. F\"olling, M. Feld, U. Schnorrberger, A. M. Rey,
         A. Polkovnikov, E. A. Demler, M. D. Lukin, I. Bloch,
         Science {\textbf{319}}, 295 (2008).


\bibitem{Hewson-book} A. C. Hewson, The Kondo Problem to
         Heavy Fermions (Cambridge University Press, Cambridge, 1993).

         \bibitem{Bauer} J. Bauer, C. Salomon and E. Demler,
          Phys. Rev. Lett. {\bf 111}, 215304 (2013).

\newpage

\bibitem{KKAK-prb-15} I. Kuzmenko, T. Kuzmenko,
         Y. Avishai and K. A. Kikoin, Phys. Rev. B {\textbf{91}},
         165131 (2015); arXiv:1402.0187.

\bibitem{IK-TK-YA-GBJ-PRB-16} Igor Kuzmenko, Tetyana Kuzmenko,
         Yshai Avishai and Gyu-Boong Jo, Phys. Rev. B {\textbf{93}},
         115143 (2016); arXiv:1512.00978.

\bibitem{IK-TK-YA-GBJ-PRB-18} Igor Kuzmenko, Tetyana Kuzmenko,
         Yshai Avishai and Gyu-Boong Jo,
         Phys. Rev. B {\textbf{97}}, 075124 (2018).

\bibitem{CIR1} Ren Zhang, Deping Zhang, Yanting Cheng, Wei Chen,
         Peng Zhang, and Hui Zhai, Phys. Rev. {\bf A} 93, 043601 (2016);
         arXiv:1509.01350.

\bibitem{CIR2} Yanting Cheng, Ren Zhang, Peng Zhang, and Hui Zhai,
         Phys. Rev. {\bf A} 96, 063605 (2017); arXiv:1705.06878.

\bibitem{CIR3} Luis Riegger, Nelson Darkwah Oppong, Moritz Hfer,
         Diogo Rio Fernandes, Immanuel Bloch and Simon Flling,
         arXiv:1708.03810.

\bibitem{CIR4} Qing Ji, Ren Zhang, Xiang Zhang, Wei Zhang, arXiv:1809.00471.


\bibitem{Yb-spectrum-78} W. F. Meggers and J. L. Tech,
         J. Res. Natl. Bur. Stand. (U.S.) {\textbf{83}}, 13 (1978).

\bibitem{Electron-afinity-PhRep04} T. Andersen,
         ``Atomic negative ions: Structure, dynamics and
         collisions".
         Physics Reports {\textbf{394}}, 157 (2004).


\bibitem{vdW-Yb-PRA-08} Masaaki Kitagawa, Katsunari Enomoto,
         Kentaro Kasa, Yoshiro Takahashi, Roman Ciury{\l}o,
         Pascal Naidon, and Paul S. Julienne, Phys. Rev. A {\textbf{77}},
         012719 (2008); arXiv:0708.0752.

\bibitem{Scatt-Length-WKB-PRA93} G. F. Gribakin and
         V. V. Flambaum, Phys. Rev. A {\textbf{48}}, 546 (1993).

\bibitem{Scazza-Yb-3P0} F. Scazza, C. Hofrichter, M. Fofer, P. C. De Groot,
I. Bloch, and S. Folling, Nature Physics {\bf 10}, 779 (2014); {\it ibid}: correction notice, (2015).


\bibitem{Scazza-Yb-3P0-2015} M. H\"ofer, L. Riegger,  F. Scazza, C. Hofrichter,
         D. R. Fernandes, M. M. Parish, J. Levinsen, I. Bloch, and S. F\"olling,
         Phys. Rev. Lett. {\textbf{115}}, 265302 (2015).

\bibitem{vdW-Yb-PRA-14} S. G. Porsev, M. S. Safronova, A. Derevianko, and
         C. W. Clark, Phys. Rev. A {\textbf{89}}, 012711 (2014).

\bibitem{Landau-Lifshitz-2} L. D. Landau and E. M. Lifshitz, The Classical Theory of Fields,
         Course of Theoretical Physics, Volume 2. (Pergamon Press, 1971).
         pp. 44-46.

\end{thebibliography}
\end{document}